\documentclass[letterpaper, 10 pt, conference]{ieeeconf}  
\IEEEoverridecommandlockouts                             
\usepackage{graphicx}      
\usepackage[table,xcdraw]{xcolor}
\usepackage{amsmath}      
\usepackage{amsfonts}      
\usepackage{algorithm}     

\usepackage{placeins}
\usepackage{siunitx}
\usepackage{textpos}
\usepackage{booktabs}
\usepackage{footmisc}

\usepackage[isbn=false,
url=true,
doi=false,
backend=biber,
style=ieee,
bibstyle=ieee, 
citestyle=numeric-comp, 
sortcites=true, 
maxbibnames=10,
giveninits=true, 
backref=false]{biblatex}
\usepackage{tikz}
\usepackage{tablefootnote}
\newcommand{\argmin}[1]{\mathrm{arg}\min_{#1}} 

\newcommand{\params}{\boldsymbol{\theta}} 
\newcommand{\tmax}{T_{\mathrm{max}}}
\newcommand{\tk}{T_k}
\newcommand{\jpartsum}{J_{\mathrm{part},k}}

\usepackage{algorithm,algpseudocode}
\makeatletter
\def\plist@algorithm{Alg.\space}
\makeatother

\usepackage{tikz}
\usepackage{textcomp}
\usepackage{lipsum}

\newcommand\copyrighttext{%
  \footnotesize \textcopyright 2024 IEEE. Personal use of this material is permitted.
  Permission from IEEE must be obtained for all other uses, in any current or future
  media, including reprinting/republishing this material for advertising or promotional
  purposes, creating new collective works, for resale or redistribution to servers or
  lists, or reuse of any copyrighted component of this work in other works.}
\newcommand\copyrightnotice{%
\begin{tikzpicture}[remember picture,overlay]
\node[anchor=south,yshift=10pt] at (current page.south) {\parbox{\dimexpr\textwidth-\fboxsep-\fboxrule\relax}{\copyrighttext}};
\end{tikzpicture}%
}

\title{\LARGE \bf
Early Stopping Bayesian Optimization for Controller Tuning
} 

\author{David Stenger$^1$, Dominik Scheurenberg$^2$, Heike Vallery$^{2,3}$, and Sebastian Trimpe$^1$
\thanks{$^1$Institute for Data Science in Mechanical Engineering, RWTH Aachen University, Germany, e-mail: \{david.stenger, trimpe\}@dsme.rwth-aachen.de}
\thanks{$^2$Institute of Automatic Control, RWTH Aachen University, Germany, e-mail: \{d.scheurenberg,h.vallery\}@irt.rwth-aachen.de}%
\thanks{$^3$Heike Vallery is also with the Department of BioMechanical Engineering, Delft University of Technology, and with the Department for Rehabilitation Medicine, Erasmus MC, Rotterdam, The Netherlands.}%
\thanks{This work was partly funded by the German Federal Ministry for Economic Affairs and Climate Action (BMWK) through the project EEMotion and by the Deutsche Forschungsgemeinschaft (DFG, German Research Foundation) under Germany's Excellence Strategy -– EXC-2023 Internet of Production -– 390621612.}
}

\usepackage[hidelinks]{hyperref}
\begin{document}

\maketitle
\copyrightnotice
\thispagestyle{empty}
\pagestyle{empty}

\begin{abstract}

Manual tuning of performance-critical controller parameters can be tedious and sub-optimal. Bayesian Optimization (BO) is an increasingly popular practical alternative to automatically optimize controller parameters from few experiments. Standard BO practice is to evaluate the closed-loop performance of parameters proposed during optimization on an episode with a fixed length. However, fixed-length episodes can be wasteful. 
For example, continuing an episode where already the start shows undesirable behavior such as strong oscillations seems pointless.
   
Therefore, we propose a BO method that stops an episode early if suboptimality 
becomes apparent before an episode is completed. Such early stopping 
results in partial observations of the controller's performance, which cannot directly be included in standard BO. We propose three heuristics to facilitate partially
observed episodes in BO.  
Through five numerical and one hardware experiment, we demonstrate that early stopping BO can substantially reduce the time needed for optimization. 
\end{abstract}

\section{INTRODUCTION}

High-level control objectives, such as energy efficiency, passenger comfort, or product quality, cannot be directly encoded in many controllers. Therefore, appropriately tuning controller parameters is essential for achieving optimal closed-loop performance. However, manual tuning can be tedious and is potentially sub-optimal. In contrast, Bayesian Optimization (BO) \cite{garnett_bayesoptbook_2023} can automatically 
optimize controller parameters for almost arbitrary objectives and controller structures on real hardware. Examples include tuning linear quadratic regulators (LQR) in robotics \cite{Marco.2016}, optimizing suspension controllers \cite{SAVAIA2021104826}, and tuning of learning-based model predictive control (MPC) in autonomous racing \cite{9981780}.

BO has been shown to be more sample efficient, i.e., it can reach a better solution with fewer experiments than metaheuristics \cite{stenger2022benchmark}. Furthermore, many advanced BO variants address important control challenges, such as safe tuning or online adaptation (see Sec.~\ref{sec:RelatedWork}).

At its core, BO leverages a probabilistic model of the analytically unknown objective function to suggest the next candidate controller, which is then evaluated in an experiment.
The vast majority of BO for controller tuning conduct experiments of \emph{fixed} episode length (see Sec.~\ref{sec:RelatedWork}).   
This common practice is, however, wasteful in terms of experimentation time, unnecessary wear, or energy consumption (e.g., \cite{kim2019bayesian}). Intuitively, an experimental episode can be stopped early if sub-optimality, e.g., oscillating behavior, of a given controller becomes apparent at the beginning of an episode.

We propose Early Stopping BO (ESBO) for time-integrated objectives such as energy consumption or mean-squared error (MSE). Based on past experiments, an application-independent rule is formulated to decide when to stop an episode. The early stopping of an episode results in a partial observation of the controller's performance. This partial observation cannot be used directly in the optimization because it would corrupt the probabilistic objective function model. To address this, we propose heuristics to facilitate the partial information. 
Deciding when to stop an episode early and how to facilitate the resulting partial information is a new problem class in BO for control (see Sec.~\ref{sec:RelatedWork}).

In summary, the main contributions of this paper are:
 
\hbox{(i) Formulate} the ESBO problem for time-integrated objectives. 
(ii) Propose heuristic approaches to decide when to stop an episode and how to facilitate partially evaluated episodes in BO. 
(iii) Evaluate effectiveness in five simulation problems and one hardware experiment. 

\section{Problem Statement: Early Stopping BO} \label{sec:ProblemStatement}

In BO, the objective is to optimally choose a parameter vector $\params$ of the control law $\pi$ that calculates the next control input $u_{t+1}$ based on past inputs $u$ and outputs $y$:  
\begin{equation}
    u_{t+1} = \pi(u_{t}, \dots, u_{1},y_{t},...,y_{1},\params).
\end{equation} 
The system dynamics  
are unknown to the optimization algorithm, and we make no assumptions about them.
The cost function $J(\params)$ captures the high-level control objective. Differently from vanilla BO, i.e., BO without early stopping (see, e.g., \cite{garnett_bayesoptbook_2023,Marco.2016,SAVAIA2021104826,charkabarty2022extermum,9981780} and Sec. \ref{sec:BO_background}), we assume that the objective is a sum of non-negative costs $j(u_{t},y_{t})$ generated at each time step $t$ of an episode with length $\tmax$: 
\begin{equation} \label{eq:objSum}
    J(\params) = \sum_{t = 1}^{\tmax} j(u_{t},y_{t}), \quad  j(u_{t},y_{t})  \geq 0.
\end{equation}
Examples of \eqref{eq:objSum} are common control engineering metrics such as the Mean Squared/Absolute Error (MSE/MAE), rise time, LQR-costs, and the integral of time-multiplied absolute value of error (ITAE). Other metrics, e.g., settling time or metrics in the frequency domain, cannot be formulated this way. 

The closed-loop behavior, i.e., $u_t$ and $y_t$, depends on $\params$, so does the cost $J$. Hence, for a known feasible set $\Theta$ of dimension $d$, the tuning problem can be formalized as: 
\begin{equation}
	\begin{aligned} \label{eq:optProblem} 
		\params^* =  \argmin{\params \in \Theta \subset \mathbb{R}^d} \qquad &   J(\params). \\
	\end{aligned} 
\end{equation}
The objective $J(\params)$ is treated as a black-box function where the objective function's analytical structure, e.g., gradients, is unknown. Thus, to solve \eqref{eq:optProblem} we sequentially take data, i.e., select parameters $\params_k$ and perform an experiment to obtain a noisy evaluation $J_k$ of $J(\params_k)$: 
\begin{equation} \label{eq:observation}
    J_k = J(\params_k) + \epsilon_k. 
\end{equation}
In almost all BO controller tuning strategies, the experiment is performed over a fixed episode length $\tmax$, data $(u_{0,k},....,y_{\tmax,k})$ is obtained and $J_k$ is calculated directly according to \eqref{eq:objSum}. The additive noise $\epsilon$ may then represent measurement noise on $y_t$ and is modeled as a Gaussian.

The objective of this work is to depart from the common assumption of a fixed evaluation length. Specifically, we consider the following problems:
\begin{enumerate} 
\item[(i)] While performing the $k$-th evaluation, at which time $\tk$ should the experiment be stopped?
\item[(ii)] If BO stopped an experiment early ($\tk < \tmax$), only data $(u_{0,k},....,y_{\tk,k})$ is available, and we cannot calculate $J_k$ directly. How can BO process the incomplete data?
\end{enumerate}

\section{Related Work} \label{sec:RelatedWork}

Bayesian optimization has successfully been applied to various controller tuning contexts \cite{Marco.2016,SAVAIA2021104826,9981780, charkabarty2022extermum,stenger2022benchmark} 
and other engineering domains \cite{garnett_bayesoptbook_2023,Shahriari.2016}. Furthermore, it can be applied to automatically optimize the hyperparameters of other learning-based control methods such as extremum-seeking control \cite{charkabarty2022extermum}, learning-based MPC \cite{9981780}, and reinforcement learning (AutoRL) \cite{parker2022automated}. 
Due to its sample-efficiency \cite{stenger2022benchmark}, the automatic tuning of controllers with BO can be attributed to the micro-data branch of reinforcement learning \cite{HandfullTries}. BO was also applied to other control engineering domains, such as state estimation \cite{NITSCH202311608} or fault diagnosis \cite{Marzat.2011}.

BO was extended to effectively address numerous advanced challenges that arise in practical control engineering applications \cite{Stenger2023thesis}, e.g., contextual optimization \cite{fiducioso2019safe}, multiple objectives \cite{makrygiorgos2022performance}, constraints \cite{khosravi2023safety}, time-varying optimization \cite{Brunzema.2022}, multi-fidelity optimization \cite{marco2017virtual}, local optimization for high dimensional problems \cite{GIBOCrash}, and safe exploration \cite{berkenkamp2021bayesian}. Shortening episodes adaptively has been used for controller tuning of an exoskeleton \cite{kim2019bayesian}. However, a domain-specific stopping criterion only valid in this specific application was used in combination with vanilla BO.  
In the controller tuning context, existing BO methods \emph{themselves} do not make active decisions about terminating an experiment. Thus early stopping is a new problem setting in BO for control. 

An early-stopped evaluation leads to partial information about the objective. 
Learning with Crash Constraints (LCC) \cite{stenger2022benchmark, Marco.2021} facilitates episodes that are not completed successfully, e.g., due to unstable behavior. However, in LCC, the decision to abort an episode is taken independently of BO, and BO only retrieves binary information on whether an evaluation was successful or not. 

Freeze-thaw BO \cite{swersky2014freeze} has been proposed to abort the training procedure of machine learning (ML) models early. It relies on two assumptions that are not given in ESBO. First, a kernel is used that relies on an exponential decay of the training loss that is typical to ML training. The additive cost \eqref{eq:objSum} does not have to follow this assumption. It does not even have to be differentiable, e.g., if the absolute error is used as a metric. Second, the model training can be resumed at any point. 
Freeze-thaw BO is extended in \cite{pmlr-v97-dai19a} by using Bayesian early stopping. However, still, an exponentially decaying kernel is used, which we cannot assume from \eqref{eq:objSum}.

Related problem formulations also occur in multi-task or robust BO. For example, \cite{Frazier.2022ExpInt} considers the problem formulation
\begin{equation}
    \max _{\params \in \Theta \subset \mathbb{R}^d} \sum_{t \in \mathcal{T}} j(\params, t) p(t),
\end{equation}
where $p(t)$ denotes the distribution of $t$. In \cite{Frazier.2022ExpInt}, BO chooses which pairs $(\params,t)$ to evaluate at each iteration. This formulation is similar to \eqref{eq:objSum} on the surface. However, the key difference is that ESBO cannot freely choose $t$. Instead, ESBO always evaluates time sequences starting with $t = 0$ up to $t = T_k$. 

In summary, to the best of our knowledge, ESBO has not been considered for controller tuning, and related methods in other domains rely on different assumptions.

\section{Background}

This section introduces Gaussian processes regression and vanilla BO without early stopping.

\subsection{Gaussian Process Regression} \label{sec:BackgroundGP}

In this paper, BO uses a Gaussian Process (GP) as a fast-to-evaluate probabilistic surrogate for the unknown scalar function $J(\params)$. 
GPs are a class of non-parametric probabilistic regression models that are defined by a prior mean function 
\begin{equation} 
m(\params)=\mathbb{E}[J(\params)]
\end{equation}
and a covariance function or kernel 
\begin{equation}
k(\params, \params')=\mathbb{E}[(J(\params)-m(\params))(J(\params')-m(\params')],
\end{equation}
where $\mathbb{E}$ is the expectation. Here, we assume $m(\params) = 0$ because we do not have prior knowledge of the objective function landscape. 

To approximate $J(\params)$, we start with a GP prior that is updated with data 
\begin{equation}
\mathcal{D}_k=\{(\params_1, J_1), ...,(\params_k, J_k)\} .
\end{equation}
This data $\mathcal{D}_k$ is obtained by performing expensive experiments \eqref{eq:observation} corrupted by identically distributed Gaussian noise $\epsilon \sim \mathcal{N}(0, \sigma_n^2)$. 
We denote the vector of noisy observations as $\boldsymbol{J}_k=[J_1, ..., J_k]^T$. 
The posterior mean $\mu_k$ and posterior variance $\sigma_k^2$ then are 
\begin{align}
    \mu_k(\params)&=\boldsymbol{k}_k(\params)^T(\mathbf{K}_k + \sigma_n^2\mathbf{I}_k)^{-1} \boldsymbol{J}_k\\
    \sigma_k^2(\params)&=k(\params, \params)-\boldsymbol{k}_k(\params)^T(\mathbf{K}_k+\sigma_n^2\mathbf{I}_k)^{-1}\boldsymbol{k}_k(\params).
\end{align} 
The gram matrix $\mathbf{K}_k \in \mathbb{R}^{k \times k}$ contains $[k(\params, \params')]_{\params, \params' \in \mathcal{D}_k}$, the vector $\boldsymbol{k}_k(\params)=[k(\params_1,\params) ... k(\params_k, \params)]$ describes the covariances between $\params$ and the observed data points $\params_1 \cdots \params_k$ and $\mathbf{I}_k$ is the $k \times k$ identity matrix. The GP model trained with data $\mathcal{D}_k$ is denoted as $\mathcal{GP}_k$. See \cite{Rasmussen.2006} for more details.

\subsection{Bayesian Optimization} \label{sec:BO_background}

Because evaluating $J(\params)$ is costly, BO leverages all data obtained until iteration $k$ to choose the next parametrization $\params_{k+1}$. After $k_\mathrm{init}$ random samples are evaluated, 
BO takes two steps to maximize the utility of the next experiment. First, a GP (see Sec.~\ref{sec:BackgroundGP}) is trained on all past observations to approximate $J(\params)$. 
Second, this model is used in an acquisition function to balance exploration and exploitation. The acquisition function $\alpha$ uses the probabilistic GP predictions to calculate the utility of an experiment. It is maximized 
to find the next query: 
\begin{equation} \label{eq:accFunArgMax}
    \params_{k+1} = \textrm{argmax}_{\Theta}
\alpha (\mathcal{GP}_k).
\end{equation}
Approximately solving \eqref{eq:accFunArgMax} is much easier than the original problem \eqref{eq:optProblem} because only the fast-to-evaluate GP model needs to be evaluated.
This new parametrization is evaluated,  
new data is received, and the next iteration is started by again updating the GP model. This way, the GP model is iteratively refined around the optimum.   
 We refer to \cite{garnett_bayesoptbook_2023,Shahriari.2016} for a detailed introduction to BO and to Sec. \ref{sec:implementation} for implementation details.     

\begin{figure*}
    \centering
    \includegraphics[width=0.95\textwidth]{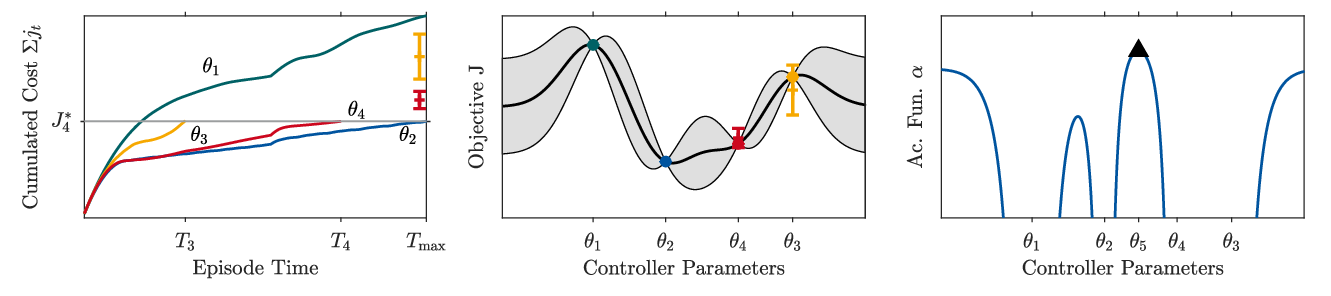}
    \caption{Illustrative example of ESBO-GP: The first and the second samples $\theta_1$ and $\theta_2$ were evaluated for the whole episode.
    The second sample $\theta_2$ achieved an improvement. The evaluation of the third and fourth samples $\theta_3$ and $\theta_4$ was stopped early, because already after $T_3$ and $T_4$, an improvement was ruled out by the early stopping rule (see Sec.~\ref{sec:StoppingRule}). 
    The objective function values of $\theta_3$ and $\theta_4$ are sampled from a probabilistic estimate 
    and included in the GP (see Sec.~\ref{sec:virtualDataSet}). The GP model of the objective function uses the observed total cost for the complete evaluations. By using the partial information in the GP, the acquisition function is smaller in areas where partial observations have been made, and the parametrization for the next episode is found by maximizing the acquisition function. }
    \label{fig:methodShort}
\end{figure*}

\section{Early Stopping BO} \label{sec:Methods}

This section extends vanilla BO to ESBO. Sections \ref{sec:StoppingRule} and \ref{sec:virtualDataSet} address the questions of when to abort an episode and how to facilitate partial observations (see problems (i) and (ii) of Sec.~\ref{sec:ProblemStatement}). The expected strengths and weaknesses of the proposed approaches are discussed in Sec.~\ref{sec:discussion}, and an illustrative example is given in Fig.~\ref{fig:methodShort}. Algorithm \ref{algo:ESBO} summarizes the proposed ESBO algorithm, and the changes to vanilla BO occur in lines 7 and 10.

\begin{algorithm}[t]
\small
\caption{Early Stopping BO}\label{algo:ESBO}
\begin{algorithmic}[1]
    \State \textbf{Input}: randomly chosen initial parameters $\params_1, \dots, \params_{k_\mathrm{init}}$ 
    \State $\mathcal{D}_k \leftarrow$ evaluate complete episodes with $\params_1, \dots, \params_{k_\mathrm{init}}$ 
    \State $k \leftarrow k_\mathrm{init}$
    \Repeat
    \State $k \leftarrow k+1$ 
    \State $J^*_k \leftarrow$ identify current optimum from $\mathcal{D}_k$ 
    \State $ \mathcal{\hat D}_k \leftarrow$ build virtual dataset using $\mathcal{D}_k$   \Comment{cf. Sec.~\ref{sec:virtualDataSet}} 
    \State $\mathcal{GP}_k \leftarrow $ fit GP model of $J(\params)$ using $\hat{\mathcal{D}}_k$
    \State $\params_{k+1} = \textrm{argmax}_{\Theta}
    \alpha (\mathcal{GP}_k)$ 
    \State $ (T_{k+1}, j_{k+1,1},\dots,j_{k+1,T_{k+1}} )  \leftarrow$ run episode with $\params_{k+1}$ and $J^*_k$ \Comment{cf. Alg.~\ref{algo:Episode}} 
    \State $\mathcal{D}_{k+1} \leftarrow \mathcal{D}_{k}  \cup \{(\params_{k+1}, T_{k+1}, j_{k+1,1},\dots,j_{k+1,T_{k+1}})\}$ 
    \Until{ $k+1 \geq K$ or $\sum T_1 \cdots T_{k+1} \geq T_{\mathrm{budget}}$  } 
    \State \Return $\params^*$
\end{algorithmic}
\end{algorithm}

\subsection{When to stop an episode?} \label{sec:StoppingRule}

Algorithm \ref{algo:Episode} states how a parametrization is evaluated in ESBO. At each time step of the episode, input and output are retrieved, and the time step-specific cost is calculated.

We choose to stop an episode at time $T_k$ if the parametrization cannot improve the current optimum, i.e., cannot return a smaller cost than the current optimum $ J^*_{k}$. From now on, $j_{t,k}$ is used to denote a sample of $j(u_k,y_k)$. Because $j_{\cdot,\cdot}  > 0$, we formulate the stopping decision rule $S$:
\begin{equation} \label{eq:stoppingRUle}
S(j_{1,k},\dots, j_{t,k}, J^*_k) :=  \begin{cases}
\mathrm{true}, & \mathrm{if} \quad \sum_{\tau = 1}^{t} j_{\tau,k} \geq J^*_k  \\
\mathrm{false},& \mathrm{otherwise}.
\end{cases}
\end{equation}
After the episode is stopped, the partially obtained cost and the stopping time are returned to BO (Line 10, Alg. \ref{algo:ESBO}). If the maximum episode length $\tmax$ is reached, the episode is also stopped, and a new best solution is found.

\begin{algorithm}[t]
\small
\caption{Episode with Early Stopping}\label{algo:Episode}
\begin{algorithmic}[1]
    \State \textbf{Input}: parameter $\params_k$, maximum episode length, $\tmax$, and current best observation $J^*_k$  
   \State Assign parameters $\params_k$ to controller and start the episode
    \For{$t = (1,2,\cdots)$} \do{ } 
    \State $(u_{t,k},y_{t,k}) \leftarrow $ obtain inputs and outputs from process
    \State $j_{t,k} \leftarrow j(u_{t,k},y_{t,k})$ calculate cost at timestep t
    \If{$t == \tmax$ or $S(j_{1,k},\dots, j_{t,k}, J^*_k)$} \Comment{cf. Sec.~\ref{sec:StoppingRule}}  
        \State 
        \Return $(T_k = t,j_{1,k},\dots,j_{t,k})$
     \EndIf
    \EndFor    
\end{algorithmic}
\end{algorithm}

\subsection{How to facilitate partial observations?} \label{sec:virtualDataSet}

If an episode is aborted early ($\tk < \tmax$), only partial knowledge $\jpartsum = \sum_{t = 1}^{\tk} j_{t,k}$ about the objective is available.
We cannot directly use $\jpartsum$ in the GP model for $J(\params)$ because it would underestimate $J(\params)$.  
Therefore, we create virtual data points that (i) guide the optimization away from the sub-optimal regions while (ii) not contradicting the smoothness assumptions of the GP and (iii) leverage the partial information. Below, three heuristics are proposed. Note that all heuristics recalculate all virtual data points at each iteration.

\paragraph{ESBO-C}
This approach treats partial observations as crashes (cf. Sec.~\ref{sec:RelatedWork}). Virtual data points are generated for all incomplete episodes $l$ by using pessimistic GP predictions as proposed in \cite{stenger2022benchmark,Stenger2023thesis}:
\begin{equation} \label{eq:ESBO-C}
    \hat{J}_l = \mathrm{min} \{ \mathrm{max}\{\bar{\mu}_J(\params_l),J^*_k\} + 3 \bar{\sigma}_J(\params_l),J_{\mathrm{max},k} \}, 
\end{equation}
where $\bar{\mu}_J$ and $\bar{\sigma}_J$ are the predicted mean and standard deviation from a GP built with only data from complete episodes. 
The objectives of the best and worst complete episodes obtained so far are denoted by $J^*_k$ and $J_{\mathrm{max},k}$. The lower bound is enforced to ensure that the virtual data point is worse than the current optimum. The upper bound constrains it in case of large predictive uncertainties.   

Because the virtual data points use probabilistic GP predictions, they ensure a smooth transition of the predictions between complete and partially observed costs. The pessimistic realization ensures that the optimization is driven away from the partial observation (cf. \cite[Fig.~4.9]{Stenger2023thesis}).

\paragraph{ESBO-TR}

The main idea of the time reformulation (ESBO-TR) approach is that with an objective function structure of \eqref{eq:objSum}, the time $T^*_m$ needed until the cumulative cost of parametrization $m$ reaches the current optimum
\begin{equation}
    \sum_{t=1}^{T^*_m} j_{t,m}  \leq J_k^*. 
\end{equation}
is a notion of fitness in itself. Thus, searching for a parametrization that would increase the episode length, i.e., that would not be aborted under the current optimum, coincides with searching for parameters with a good objective function value. 
At each iteration $k$ for all observations $m \in \{1,\dots, k\}$, we find the time $T^*_m$, after which the cumulative cost exceeds the current observed optimum $J^*_k$  
and assign it as the virtual objective function value: $\hat{J}_m  = -T^*_m$. Thus, for the current optimum, we assign $-T^*_\mathrm{max}$. The minus sign is needed because we maximize episode length to minimize cost.

\paragraph{ESBO-GP} \label{sec:ESBO-GP}

The main idea is to probabilistically estimate the unobserved portion, $T_l$ to $\tmax$, of partially observed episode $l$ with separate GPs (see Fig. \ref{fig:methodShort}).  

First, the time period of an episode is split into different sections, the first section starting at $T^{\mathrm{sort}}_1$ and ending at $T^{\mathrm{sort}}_2$: 
\begin{equation*} \label{eq:split}
    (T^{\mathrm{sort}}_1,\dots,T^{\mathrm{sort}}_{N_{\mathrm{GP}}+1}) = \mathrm{unique}(T_1,\cdots T_{k}).
\end{equation*}

The unique operator sorts the previously obtained episode lengths and removes duplicates. In the example of Fig. \ref{fig:methodShort}, the unique episode lengths are $(T_3,T_4,T_\mathrm{max})$. Next, $N_{\mathrm{GP}}$ GPs are built to model the cumulated cost $j_{\mathrm{sec},\cdot}$ for each section between the unique episode lengths:    
\begin{equation} \label{eq:partialGPs}
    \begin{aligned}
    j_{\mathrm{sec},1}(\params) = \textstyle\sum_{t=T^{\mathrm{sort}}_1}^{T^{\mathrm{sort}}_2} j (u_t,y_t) &\sim \mathcal{GP}_{1}, \\ 
    \cdots & \\
    j_{\mathrm{sec},N_{\mathrm{GP}}}(\params) =  \textstyle\sum_{t=T^{\mathrm{sort}}_{N_{\mathrm{GP}}}}^{T^{\mathrm{sort}}_{N_{\mathrm{GP}}+1}} j (u_t,y_t)  &\sim \mathcal{GP}_{N_{\mathrm{GP}}}.  
    \end{aligned}
\end{equation}
The GPs are trained using all data available for the respective sections. In the example of Fig. \ref{fig:methodShort}, two additional GPs are trained, one from $T_3$ to $T_4$ and one from $T_4$ to $T_\mathrm{max}$. 
Using the GP models, we make probabilistic predictions for each section $q$ and partially evaluated parametrization $l$:
\begin{equation}
    j_{\mathrm{sec},q}(\params_l)  \sim \mathcal{N}(\mu_q(\params_l) ,\sigma^2_q(\params_l)), \forall q \in \{1, \cdots N_{\mathrm{GP}} \} .
\end{equation}
These are then added to the partial evaluation to generate the probabilistic predictions for the full cost:
\begin{equation}
\begin{aligned}
    \tilde{J}_l  & \sim \mathcal{N}(\mu_{\tilde{J}_l},\sigma^2_{\tilde{J}_l}), \quad \mathrm{with}\\
    \mu_{\tilde{J}_l}  & = \sum_{t = 1}^{T_{l}} j_{t,l} + \mu_p(\params_l) + \dots + \mu_{N_{\mathrm{GP}}}(\params_l) \\
    \sigma^2_{\tilde{J}_l} & = \sigma^2_p(\params_l) + \dots + \sigma^2_{N_{\mathrm{GP}}}(\params_l).  \\
\end{aligned}
\end{equation}

We cumulate the probabilistic estimates starting with GP model $p$ that models the section starting at $T_l$. 

The virtual data point $\hat{J}_l$ is sampled randomly at each iteration from the distribution of $\mathcal{N}(\mu_{\tilde{J}_l},\sigma^2_{\tilde{J}_l})$. Additionally, we bound $\hat{J}_l$ to be larger than the known partially obtained cost $\sum_{t = 1}^{T_{l}} j_{t,l}$. 

We also tried using the probabilistic estimate $\tilde{J}_l$ directly in the GP for $J(\params)$ as a virtual data point with known heteroscedastic noise or model $j(\params,t)$ directly as a GP similarly to \cite{swersky2014freeze}. These two options were not pursued further because preliminary results were not promising. The former option can lead to a large predictive uncertainty at early stopped parametrizations resulting in a non-zero acquisition function. However, this uncertainty cannot be reduced by resampling the same location.

\subsection{Expected Strengths and Weaknesses} \label{sec:discussion}

In BO, the evaluation of a parametrization serves two main purposes. First, we want to determine if the parametrization leads to a better solution than previously found. This is ensured by \eqref{eq:stoppingRUle}. Secondly, completing an episode may be valuable in improving the GP model to make better-informed future parameter choices, especially if substantial time is needed to reset the episode. The latter is not considered here. In contrast, an episode could be aborted even earlier by using a probabilistic model of the future cost of the remaining episode \cite{swersky2014freeze} (see also Bayesian early stopping \cite{pmlr-v97-dai19a}). This may be especially useful in case the majority of the cost is generated late in the episode and is also not considered here.  

It is expected that using more data from partial observations is beneficial for the progress of the optimization. ESBO-C only uses the binary information on whether the episode was stopped early. ESBO-TR uses information until the respective episodes would have been stopped under the current optimum, and ESBO-GP leverages all available data from the partial observations.  

Noisy optimization may be challenging for ESBO-TR because we cannot directly model the observation noise on $J_k(\params)$ by applying the time reformulation.

\begin{figure*}
    \centering
    \includegraphics[width=0.99\textwidth]{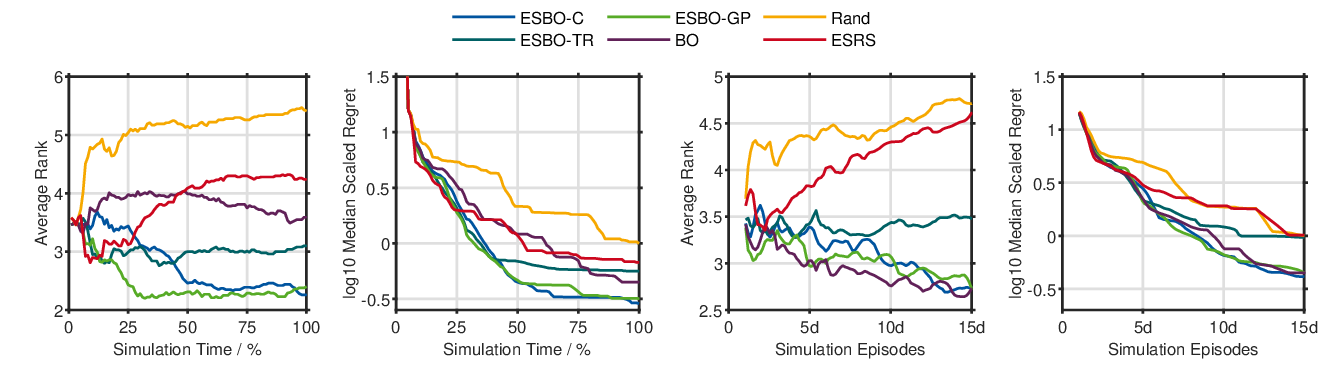}
    \caption{Results of the simulation study. For an explanation of the metrics see Sec.~\ref{sec:Metrics}. \label{fig:sim_results}}
\end{figure*}

\section{Empirical Evaluation} \label{sec:results}

After introducing hyperparameters and evaluation metrics, this section evaluates the proposed ESBO variants on five deterministic stimulation and one hardware controller tuning tasks. Our experiments reveal the following key findings:

\begin{itemize}
    \item In simulations, ESBO-C and ESBO-GP perform similarly and outperform ESBO-TR, vanilla BO, and random search baselines. 
    \item In simulations, ESBO-GP speeds up optimization by up to $48 \%$ simulation time steps in comparison to vanilla BO without compromising the quality of the solution at maximum budget. 
    \item In the hardware experiment,  
    ESBO-GP speeds up optimization by up to $35 \%$ experimentation time and finds final solutions comparable to vanilla BO. 
\end{itemize}

Our implementation used in the experiments is available\footnote{\label{note1} \href{https://github.com/Data-Science-in-Mechanical-Engineering/ESBO}{https://github.com/Data-Science-in-Mechanical-Engineering/ESBO}}.

\subsection{Implementation and Hyperparameters}  \label{sec:implementation}

The simulation and hardware experiments aim at determinining the of-the-box-performance of the proposed algorithms. Therefore, none of the BO hyperparameters were adjusted to the test cases, and all experiments use identical hyperparameters.

Max-value entropy search (MES) \cite{wang2017max} is used as the acquisition function. MES does not rely on additional decisive parameters and applies to noisy and deterministic optimization. 
The acquisition function is multimodal. Therefore, the acquisition function maximization \eqref{eq:accFunArgMax} is performed by a combination of random sampling and gradient descent. 

For GP modeling, the input $\params_1,\dots,\params_k $ is transformed to the unit cube, and the output $\hat{J}_1,\dots,\hat{J}_k$ is normalized to zero mean and a standard deviation of one. We use a zero prior mean and an anisotropic squared-exponential kernel for the transformed data. 
GP-hyperparameters (signal strength, length scales, and observation noise for the hardware experiments) are optimized in each iteration by maximizing the log-likelihood. Additionally, length scales are bounded such that a correlation of $0.1$ is possible over at most half the domain and at least over $1\%$ of the domain. This ensures that length scales are within a similar order of magnitude as the optimization domain.

In the hardware experiment and three of the five simulation experiments, some parameterizations lead to crashes due to instability or tank overflow (cf. crash constraints  Sec.~ \ref{sec:RelatedWork}). Here, we use BO-VDP \cite{Stenger2023thesis} to generate virtual data points according to \eqref{eq:ESBO-C} for the crashed evaluations. Our implementation\footref{note1} 
is based on the GPML toolbox\footnote{\href{http://gaussianprocess.org/gpml/code/matlab/doc/}{http://gaussianprocess.org/gpml/code/matlab/doc/}}.

\subsection{Evaluation Metrics} \label{sec:Metrics}

We use different test cases and seeds, i.e., initial samples, to empirically evaluate ESBO's performance. 
Simple regret and rank are used to quantify the optimization progress. The simple regret $r_k$ after $k$ evaluations is calculated as the difference $r_k = J^*_k - J^*$ between the best evaluation found so far $J^*_k$ and the overall optimum $J^*$. We take the best evaluation across all algorithms and seeds for $J^*$. To make it comparable across different algorithms, we scale the regret such that a scaled regret of $1$ corresponds to the median performance of random search at maximum budget. The scaled regret is then averaged over all test cases and seeds. \cite{stenger2022benchmark}   

We sort the regrets of the different evaluated algorithms to calculate the rank. The algorithm with the lowest regret gets ranked $1$, the second best gets ranked $2$, and so on. The average rank is calculated by simply averaging over test cases and seeds. \cite{stenger2022benchmark}

The motivation of ESBO is to reduce total interaction time, i.e., simulation time steps or hardware experimentation time. Therefore, results are reported with respect to interaction time and evaluations.

\subsection{Simulation Study} \label{sec:resSim}

\paragraph*{Setup} Table \ref{tab:testcases} summarizes the simulation test cases. Five controllers ranging from a simple bang-bang controller to a linear MPC are optimized. Plant models vary from a synthetic second-order system (PT-2) with dead time to a more complex non-linear three-tank system.  
\begin{table} 
	\caption{Simulative controller tuning tasks with problem dimension $d$}
	\centering
	\begin{tabular}{p{0.5cm}| >{\raggedleft}p{1.3cm} >{\raggedleft}p{1.3cm} >{\raggedleft}p{0.3cm} >{\raggedleft}p{1.3cm} >{\raggedleft}p{0.9cm} }
		\toprule
		  No.  & Controller & Plant   & $d$  & Objective & Crashes? (\ref{sec:implementation})                      \tabularnewline
		\midrule
		  	1 & Bang-Bang &  Boiler
     &  $1$   & Energy \& Comfort  & No  
		    \tabularnewline %
		   \rule{0pt}{3ex}%
		    2 & PI &  Three Tank \cite{Scheurenberg23} &  $2$   & MSE & No    
		    \tabularnewline %
		   \rule{0pt}{3ex}%
		   3  & PID &  PT-2 w. Dead-Time & 3 &  ITAE & Yes              
		    \tabularnewline%
        \rule{0pt}{3ex}

         4 & State Feedback  &  Cart Pole
         &  $4$              &  LQR-Costs & Yes 
		   \tabularnewline%
        \rule{0pt}{3ex}%
        5 & MPC  &  Three tank 
        \cite{Scheurenberg23}    &  $5$              &  MAE & Yes  
        		   \tabularnewline
		\bottomrule
	\end{tabular}
	\label{tab:testcases}
\end{table}

To evaluate the effectiveness of ESBO-C, ESBO-TR, and ESBO-GP, we use three baselines: (i) Random Sampling with early stopping (ESRS), i.e., choosing parameters $\params$ at random and using the early stopping rule \eqref{eq:stoppingRUle}, (ii) Random Sampling (RS), i.e., same as ESRS but without early stopping, (iii) BO without early stopping (BO). 
 
Each optimization is run with 10 different random seeds. The initial sample size is set to the problem dimensionality $k_{\mathrm{init}} = \mathrm{max}\{d,2\}$. The optimization is terminated if $T_{\mathrm{budget}} = 15 \cdot d \cdot  \tmax$ time steps or $K = 45 \cdot  d$ objective function queries are exceeded. Box constraints $\params_\mathrm{min} \leq \params \leq \params_\mathrm{max}$ define the domain $\Theta$.

\paragraph*{Results} Figure \ref{fig:sim_results} shows the simulation results averaged over all test cases. All three heuristics reach better results faster than vanilla BO and RS. The simulation time needed to achieve the median final performance of ESRS is reduced by $48 \%$ from $62 \%$ to $32 \%$ percent using ESBO-GP instead of BO. The performance after $T_\mathrm{budget}$ is best using ESBO-C and ESBO-GP. The ESBO variants need comparably many evaluations as vanilla BO and the RS variants are not competitive.

\subsection{Hardware Experiments} \label{sec:resExp}

\begin{figure}[t]
    \centering
    \includegraphics[width=0.9\columnwidth]{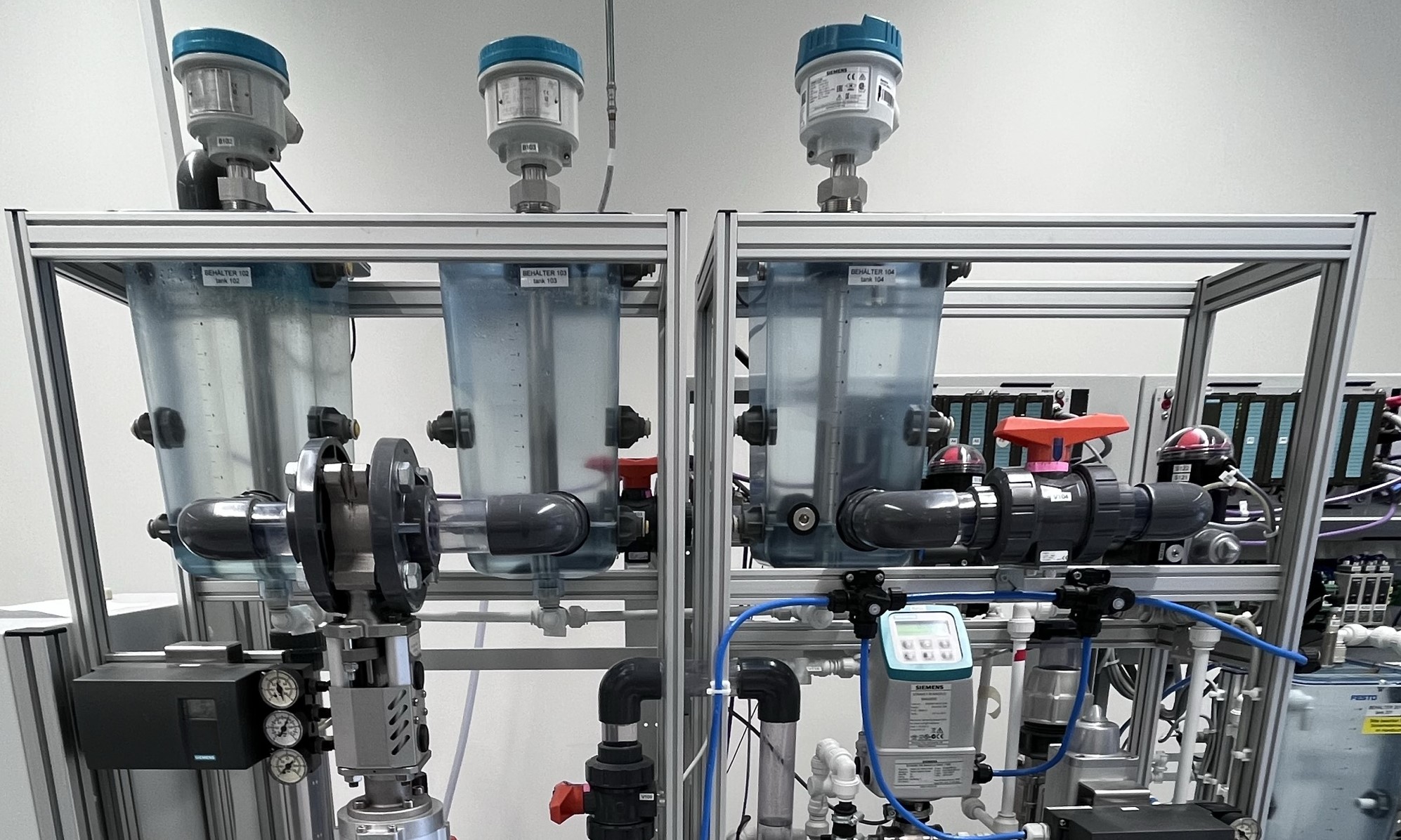}
    \caption{Three tank hardware test bed at the model factory of \emph{Institute of Automatic Control, RWTH Aachen University}.}
    \label{fig:tank}
\end{figure}

\paragraph*{Setup} We evaluate the practical applicability of the ESBO method on a three-tank hardware test bed (cf. Fig.~\ref{fig:tank}).  
A detailed description of the test bed is given in \cite{Scheurenberg23}.

A PI controller tracks a reference step in the desired water level of the third tank. Thus, the proportional and integral gains of the PI controllers are optimized in the log domain. The objective $J$ is the mean squared tracking error plus a penalty on large pump actuation.

Because it performed most consistently in simulation ESBO-GP is evaluated on the hardware test bed with BO and ESRS as benchmarks. All algorithms are run with eight randomly selected starting parameterizations each. 
The same budget and metrics as in Sec.~\ref{sec:resSim} are used, with the exception of an initial sample size of $k_\mathrm{init} = 3$. Note that in contrast to the simulation study, noise is now present. Therefore, BO and ESBO-GP include the observation noise $\sigma_n^2$ in the hyperparameter optimization.

\begin{figure}[t] 
    \centering
    \includegraphics[width = \columnwidth]{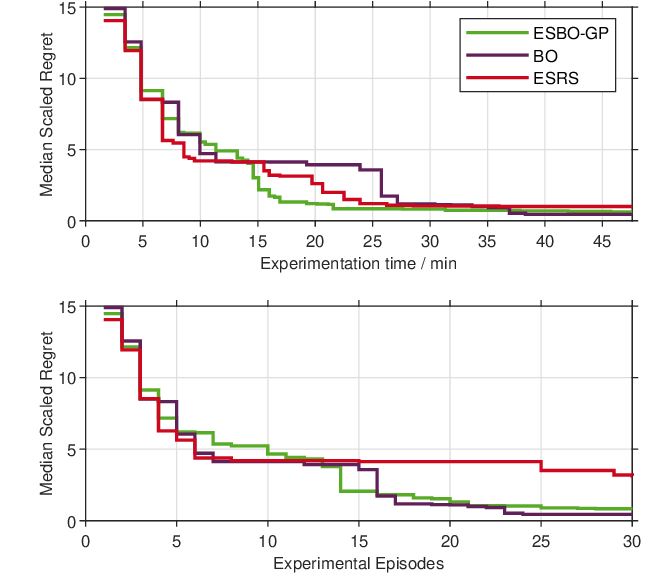}
    \caption{The two plots represent the same hardware experiments, showing median scaled regret over experimentation time (top) and number of episodes (bottom). For an explanation of the metrics see Sec.~\ref{sec:Metrics}.}
    \label{fig:ExpResTS}
\end{figure}

\paragraph*{Results} Figure \ref{fig:ExpResTS} shows the optimization progress. 
ESBO-GP reaches better solutions at around $15$ to $26$ minutes and achieves a comparable final solution. The experimentation time needed to achieve the median final performance of ESRS is reduced by $35\%$ from $34$ to $22$ minutes using ESBO-GP. 
 
Both BO variants outperform random search when considering progress w.r.t. episodes (queries).

We evaluated the statistical significance of the results using a rank-sum test at $5 \%$ significance level. After around $30$ minutes, the performance difference is statistically insignificant, while ESRS and BO perform significantly worse than ESBO-GP at around $15$ to $26$ minutes. Note that the slight differences in initial performance after experiment one occur due to process noise.   

\subsection{Discussion}

In addition to reducing interaction time, the ESBO variants and BO exhibit similar query efficiency. This indicates that the ESBO heuristics proposed to generate the virtual data points are sufficient to enable optimization although samples are less informative in ESBO. 
ESBO-GP is the best-performing heuristic possibly because it uses all information obtained from the partial evaluations. Early stopping benefits random search substantially, making it a competitive alternative to vanilla BO. Random search is outperformed consistently, highlighting that the hardware and simulation test cases chosen are not trivial to solve.  

\section{Conclusion} \label{sec:Conclusion}
We introduced ESBO to automate controller tuning, a setting where BO stops the closed-loop evaluation before nominal completion. We proposed heuristic approaches to address ESBO's key challenges: (i) decide when to stop an episode and (ii) enable the resulting partial evaluations. 

The proposed methods were evaluated in five diverse simulation tasks and one hardware experiment. Overall, results highlight the potential of ESBO to speed up automated controller tuning without sacrificing final solution quality. The developed heuristics may serve as baselines for more principled early stopping BO approaches for controller tuning in future research. 

\section*{Acknowledgments}

We thank Alexander von Rohr, Johanna Menn, and Paul Brunzema for many helpful comments, suggestions, and discussions. 

\AtNextBibliography{\footnotesize}

\printbibliography

\addtolength{\textheight}{-6cm}

\newpage

\end{document}